\documentclass[runningheads]{llncs}
\usepackage[T1]{fontenc}
\usepackage{graphicx}
\begin{document}
\title{Fault Tolerant Reconfigurable ML Multiprocessor}
%
%\titlerunning{Abbreviated paper title}
% If the paper title is too long for the running head, you can set
% an abbreviated paper title here
%
%\author{First Author\inst{1}\orcidID{0000-1111-2222-3333} \and
%Second Author\inst{2,3}\orcidID{1111-2222-3333-4444} \and
%Third Author\inst{3}\orcidID{2222--3333-4444-5555}}
\author{Tangrui Li \inst{1}
\email{tuo90515@temple.edu}
\orcidID{0009-0005-3471-038} \and
Justin Y. Shi \inst{1}
\email{shi@temple.edu}
\orcidID{0000-0003-3774-2190} \and
Matteo Spatola \inst{1}
\email{matteo.spatola@temple.edu} \and
%\orcid{Your ID}
Hongzheng Wang \inst{1}
\email{hongzheng.wang@temple.edu}
%%\orcid{Your ID}
}

\authorrunning{T. Li \& J. Shi et al.}
% First names are abbreviated in the running head.
% If there are more than two authors, 'et al.' is used.
%
%\institute{Princeton University, Princeton NJ 08544, USA \and
%Springer Heidelberg, Tiergartenstr. 17, 69121 Heidelberg, Germany
%\email{lncs@springer.com}\\
%\url{http://www.springer.com/gp/computer-science/lncs} \and
%ABC Institute, Rupert-Karls-University Heidelberg, Heidelberg, Germany\\
%\email{\{abc,lncs\}@uni-heidelberg.de}}
%

\institute{Temple University, Philadelphia PA 19102, USA}
\maketitle              % typeset the header of the contribution
\begin{abstract}
This paper reports three computational experiments for a von Neumann inspired reconfigurable fault tolerant multiprocessor for neural network (NN) training workflows. The experiments are intended to prove the feasibility of the proposed reconfigurable multiprocessor architecture for non-regular workflows on robustness of adaptability. A potential integration with MLIR compilers is also discussed for integrating diverse accelerator hardware for existing practical applications.
\end{abstract}
\section{Introduction}

The rapid growth of NN models has led to a surge in computational demands. While high-performance GPU clusters seem to meet these needs, such devices are becoming cost-prohibitive due to thermal management difficulties and checkpoint overheads. These costs grow as infrastructure expands. A renewed research interest has shifted toward utilizing multiple devices rather than centralized GPU clusters. However, scaling the NN models across multiple heterogeneous devices introduces three major challenges:  1) \textbf{Single Point of Failure (SPF)}: As the system grows, the probability of a single device failure bringing down the entire application increases. 2) \textbf{Inter-device Communication Overhead}: Frequent data transfers between devices introduce substantial latency and bandwidth pressure, especially as model size and number of devices increase. 3) \textbf{Load Balancing}: Achieving high memory utilization and compute efficiency requires partitioning of the computation graph to ensure prompt termination without processors idling at the same time.

Eliminating SPF in multiprocessor clusters has been an open challenge. Even the latest GPU clusters require frequent checkpoints. The checkpoint time and storage overheads grow as the processing infrastructure expands. Checkpoint is like a "shotgun approach" to defend against potential loss of computational results. Creating reconfigurable multiprocessors is another open challenge that requires mapping arbitrary workflows (NN models in this paper) to processors to ensure high processing and communication efficiencies. Recently, a von Neumann principle inspired \cite{vonNuemann} fault-tolerant reconfigurable multiprocessor was reported to have solved both challenges at the same time \cite{Shi2025}. Due to the theoretical impossibility of reliable failure detection \cite{LynchImpossibility}, all application programs are required to handle timeout/retransmission discipline as an integral part of the distributed application. A timeout event signals a potential component failure, a slow network, or a busy or slow component. A simple retransmission logic with redundancy elimination can solve the application fault tolerance challenge without explicit checkpoints. Carefully developed the timeout/retransmission discipline can result in significant load-balancing savings for non-regular workflows by tuning task distribution parameters. 

This paper reports a computational simulation result for such a fault-tolerant reconfigurable processor for NN training applications with a focus on robust load balancing over heterogeneous processors. 
% 关于SPF的部分你只需要找一些体现出SPF问题的神经网络并行训练方法就可以了
% 关于device communication overhead的部分我觉得也很常见，只不过在一般的文献里都不会主动描述自己的缺点，你可以找在一般的通信系统里由多设备引起的overhead，比较典型的例子就是为什么不能使用低质量设备海洋替代高性能数据中心
% 关于Manual Partitioning的东西和下面有关MP的部分是重合的，我们在下面说
% 关于异构（heterogeneous）设备的出现，你可以参考最近有关于TPU的文献，它们一般为了介绍自己的设备往往会涉及自己的TPU与现在的计算设备的对比，这种设备往往有很多，比如tenstorrent

The fault tolerant reconfigurable multiprocessor architecture is enabled by the ACAN data representation and program-program synchronization semantics in order to map arbitrary workflows to any multiprocessor cluster or clusters. This framework naturally compliments traditional optimizing compiler efforts, such as MLIR (Multi-Level Intermediate Representation) \cite{10.1109/CGO51591.2021.9370308} programs generated by the compilers assume unrealistic component reliabilities.  The compilers provide optimized program mapping for different applications by enabling dialect-based abstraction and transformation on specific accelerator hardware. Optimized program mapping at the global processing graph level is difficult.  The proposed method can thus solve the broader scalability challenge of removing significant barriers to the adoption of MLIR-based compilers in large-scale AI/ML projects.

% 我相信有很多关于MLIR的文献

While a complete system would involve systematic dialect specification and backend-specific lowering, this work focuses on a proof-of-concept implementation using a simplified NN class composed of linear layers only. At this stage, we do not provide a full dialect design or device lowering pipelines, as those require extensive engineering efforts. Instead, our approach introduces a hardware-agnostic abstraction layer via the TS model, which can conceptually align with any MLIR-compatible NN and be lowered flexibly in future accelerator implementations.

The remainder of this paper is organized as follows. We first review related work on parallel training of neural networks, followed by an introduction to the key components of our system: Tuple Space and Active Content Addressable Networking. Then, we detail how these components interact to support efficient NN parallelization, and present experiments evaluating feasibility, adaptability, and robustness. A discussion on MLIR integration and potential lowering strategies is also included. 

\section{Related Work}

\textbf{\indent Data Parallelism (DP)} 

% DP的核心你可以参考下面的描述，简单来说就是每个设备都有完全相同的模型，但是跑的数据不一样，最后需要某一个设备上聚合起来

The most straightforward distributed training strategy, where each device holds a full copy of the model and processes distinct data batches. Gradients are computed locally and must be aggregated on a primary device \cite{NIPS2012_6aca9700,Horovod}. This synchronization becomes a scalability bottleneck. Moreover, memory utilization is inefficient, as all devices redundantly store the same model copy.

\textbf{Distributed Data Parallelism (DDP)}

% DDP 是为了解决DP最后“需要在某一个设备上集中”的问题而产生的，具体的可以看下面的描述。

To mitigate the communication bottleneck of DP, DDP\cite{li2020pytorchdistributedexperiencesaccelerating} was proposed to decentralize gradient aggregation. The most widely used strategy is Ring AllReduce \cite{10.1016/j.jpdc.2008.09.002,gibiansky2017bringing}, where each device exchanges partial gradient tensors with its neighbors in a ring network topology. Over several rounds of communication, all devices converge to the same gradients. High-performance libraries like Nvidia NCCL\footnote{https://github.com/NVIDIA/nccl} and accelerator hardware like Tenstorrent\footnote{https://github.com/tenstorrent/} can satisfy such requirements.

% ring all-reduce是一个比较有名的DDP算法
% NCCL是nvidia的一个包，相信很简单就能找到
% 需要一些介绍tenstorrent硬件的文献，只要是提到它们有很多processor，而且processor之间的通信能力的都行

\textbf{Model Parallelism (MP)}

Unlike DP focus on data, MP\cite{10.5555/2999134.2999271} distributes the model across devices. This is often necessary when the model is too large to fit into a single device. MP is generally implemented in two main categories. 

\textbf{Pipeline Parallelism (PP)} By partitioning modules sequentially across devices, pipeline parallelism enables concurrent processing of multiple micro-batches in different stages\cite{10.5555/3454287.3454297,10.1145/3341301.3359646}. However, pipeline bubbles and imbalances between stages can lead to low resource utilization. On the other hand, such pipelines are usually manually designed, although recent approaches, such as Alpa \cite{280874}, enable automated pipeline partitioning; such methods still face limitations, such as: 1) They are primarily effective for a limited set of model architectures with sequential or layered structures. 2) The underlying combinatorial search for optimal partitions is computationally expensive. 3) When the external hardware configuration changes, the pipeline often needs to be recomputed from scratch.

% pipeline parallelism和很久之前我们和施老师讨论的“流水线”是类似的，也就是把神经网络的不同层分散在不同的设备上，已形成pipeline。它的问题主要有两个，第一个是这个分割往往不是自动的，因此这里你可以把reference添加回上面提到的manual partitioning上。第二个问题是pipeline的同步需要很精细，不然就会有设备等待之前设备的完成，因此容易出错。任何有关这上面topic的都可以。另外的pipeline parallelism可能有一个别名“inter-layer parallelism”。我推荐在这里可以多找几篇。
% Alpa 就是研究名，很简单

\textbf{Tensor Parallelism (TP)} This finer-grained strategy partitions the computation within individual layers across multiple devices. Megatron-LM \cite{shoeybi2020megatronlmtrainingmultibillionparameter,10.1145/3458817.3476209}, a widely adopted implementation of tensor parallelism, requires manual modification of the model code to integrate its custom parallel primitives. This constraint poses challenges when adapting large-scale or pre-existing codebases.

% tensor parallelism也有一个别名，就是intra-layer parallelism，从名字可以看出和上面那个是一对的，它强调的是一个层要怎么并行，一般涉及对一层的精细拆分，一般需要的是一些软件包。如果使用这些软件包的封装，那没问题，但是代价是必须修改源码使用那些软件包，我相信有不少针对这个的介绍和批评。
% 就比如下面说到的Megatron，你可以多找一点

\textbf{ZeRO (Zero Redundancy Optimizer)}

% 我相信你已经很熟悉ZeRO了

Instead of replicating all parameters, gradients, and optimizer states across devices, ZeRO\cite{10.5555/3433701.3433727} partitions these elements among devices and reconstructs only the necessary parts during computation, significantly improving memory efficiency and scalability. As implemented in DeepSpeed \cite{10.1145/3394486.3406703} through progressive stages (ZeRO-1, 2, 3), this approach trades off communication overhead for a reduced memory footprint.

However, while ZeRO excels at memory optimization, it does not alter the computation graph and can not resolve issues where individual layers exceed device memory limits. The benefit is that ZeRO does not require manual source code change, but if the original model contains a layer that exceeds the memory capacity of a single device, ZeRO will not be able to handle it.

\textbf{Guided Self-Scheduling}

It has long been recognized that the task size assigned to a processor will have a significant impact on the overall performance and efficiency of computation. Most practical application task durations are unpredictable and vary significantly. Guided Self-Scheduling (GSS) aims to automatically determine parallel tasks for all applications, minimizing idle times and maximizing processor efficiency by evaluating task response times \cite{GSS}. For non-regular applications using heterogeneous processors, GSS uses a "small->large" heuristic tuning strategy. Faster processors would complete the small tasks quickly while the slower processors became busy most of the time. Eventually, the faster processors would get more work done than slower ones automatically.

\section{Tuple Space and ACAN}

Active Content Addressable Networking (ACAN) departs from Content Addressable Networking (CAN)\cite{10.1145/964723.383072} in two key ways: 1) networked data representation, and 2) networked data access protocols.  Traditional CAN is typically implemented using a Distributed Hash Table (DHT) for quick data retrieval. The data contents are linked to "buckets" inside a DHT \cite{DHT}.  Therefore, the network data representation is <data, bucket ID>. The data accesses protocols are "read" and "write". The data-bucket ID binding violated von Neumann's  principle of reliable system synthesis \cite{vonNuemann}, since the "bucket ID" may not be available sometimes. It is the cause for single-point failures. 

ACAN's networked data representation is <key, value>. It supports three data access methods: 1) put(key, value), 2) read(\&pattern, \&buffer) and 3) get(\&pattern, \&buffer), where the read($\bullet$) and get($\bullet$) are blocking calls for synchronization between programs. These "Tuple Space" semantics are supported by all participating nodes as an extension to the TCP/IP stack. Nodes within the "Tuple Space" communicate via automatic peer-to-peer discovery protocols. Together, they form an application-aware computing cloud.

ACAN completely decouples application programs and data from physical and virtual components. The program/data decoupling from devices and the tuple space communication with synchronization semantics enable runtime systems to perform automatic fault isolation and load balancing. At runtime, the application's workflow will automatically form SIMD (Single Instruction Multiple Data), MIMD (Multiple Instruction Multiple Data) pipelines using all available resources. The following section explores this idea in-depth.

For NN training applications, ACAN can improve the performance and fault tolerance of compute. First, ACAN significantly reduces single point of failure risk by enabling run-time fault isolation and task rerouting unless all devices fail at the same time. Second, as discussed in timeout/retransmission discipline, the TS semantics enable efficient communication and synchronization between heterogeneous devices without requiring manual directions. Third, the proposed method requires the application programs to handle timeout events in support of automatic graph partitioning, eliminating the need for manual scheduling or slicing. Finally, this method enables dynamic device participation, allowing new devices to join or leave at run-time while maintaining service continuity, making the system suitable for non-deterministic environments.

\section{Using a Custom ACAN Cloud}

An ACAN application typically consists of a single manager and multiple handlers. Both the manager and the handler can modify the tuple space. For mission-critical applications, all updates can be logged in an immutable blockchain, ensuring traceability and accountability.

The manager is responsible for publishing tasks to tuple space while handlers pull tasks from it. After the manager publishes a task, each handler will get($\bullet$) the tasks that it can handle according to the content of the task and its own capabilities. A get($\bullet$) request will populate the custom cloud until a match is found. When a task is taken after a match, other handlers will no longer see it.  The handler who receives this task can choose whether to "process" or "store" the task. A stored task may be matched later by a different handler. When multiple handlers are competing for the same get() pattern, SIMD parallelism is formed automatically. When multiple handlers are asking for different patterns, MIMD will form. The natural dependencies between the input and output tuples will form task pipelines seamlessly.

In the ACAN architecture, the manager and handlers operate without direct communication, enabling the runtime to perform global resource multiplexing efficiently. Once the manager dispatches a task, a timer (timeout) is set to allow handlers to process it. Upon timeout, the system evaluates the completion status of all dispatched tasks. If a task remains uncompleted, the manager can re-send it. Similar to GSS, we initially use a small fixed task size, that is, we try to test if the processing difficulty of each task is close enough, so that the higher-performance handler can process the task and complete at approximately the same time without knowing the processor details. The ideal situation involves tasks being released at one time can be completed at the same time regardless of the performance of the handlers. Considering that the service of the handler is not necessarily reliable, the manager can adjust 1) the task size, 2) the size of the task released at one time (called pouch size), and 3) the timeout for waiting for the completion of the released task to finetune the deliverable application performance.

The custom ACAN cloud allows any handler to fail at any time. The loss of a Handler will only affect the task pouch size and timeout. The Manager has the option to store the partial result in a stable storage or leave them in TS. And the crash of Manager should not damage the partially computed results. If needed, the Manager restart can be programmed to read the tuple space state and continue. Since the Manager uses a single processor, the current industry record of single processor MTBF (Mean Time Between Failure) is more than 11 years \footnote{https://chatgpt.com/share/6876cffe-0d6c-800b-908a-e4d47d94ca4b}. Unless the application requires that long running time, the Manager checkpoints may not be necessary.

\section{Method}

This section focuses on deploying NNs using ACAN. To that end, we first introduce the deployment content and its TS representation, then we will introduce the dispatching and conflict resolution details.

\subsection{Individual ACAN Task}

Considering that the current framework focuses solely on NNs composed of linear layers, we first obtain the necessary parameters (input and output dimensions) either through manual input or by parsing the MLIR. Based on this information, three prototype task types are generated for each layer: \textbf{forward}, \textbf{backward}, and \textbf{update}. If the layer is not the last layer, an additional \textbf{activation} task is created (since we assume a regression scenario). For the last layer, a \textbf{loss} task is created instead.

Each task type has specific execution preconditions:

\begin{itemize}
    \item Forward task: Requires the output of the previous layer (or the input data, if it is the first layer) to be available in the tuple space. Otherwise, the task fails upon timeout and is discarded.
    \item Activation task: Requires the hidden variable to be activated.
    \item Loss task: Requires both the final output and the ground-truth label.
    \item Backward task: Depends on the availability of the gradient from the next layer (or the gradient derived from the loss function, if it is the last layer). This task computes the first-order derivative based on the chain rule (without involving advanced optimization techniques, though such extensions are possible).
    \item Update task: Requires the corresponding gradient for the parameter to be updated.
\end{itemize}

TS is queried using variable names as keys to determine task readiness. Although this leads to frequent accesses, the design characteristics of TS mitigate performance concerns.

Notably, when tasks are created, they are represented as declarative descriptions (e.g., a string of the task type and parameters) rather than instantiated objects. This design enhances the decoupling between the Manager and Handler in ACAN. It is important to emphasize that these are merely prototype tasks. They require further processing before being submitted to the tuple space.

Together, these tasks define the core execution pipeline for NN training. The associated execution states, such as current data items or active layers, are also recorded in TS, serving as the foundation for task dispatching, execution, and validation by the Manager and Handler.

\subsection{Prototype Task Partition}

As discussed earlier, all tasks are of the same size for ease of tuning to achieve optimal utilization of system computational resources in a Handler-agnostic manner. This allows for consistent control of execution granularity through the adjustment of pouch size and timeout. However, since NN layers vary in size, the prototype tasks derived from the network structure must be partitioned to satisfy the above constraints.

Since tasks are represented as descriptions rather than instantiated objects, the designated size can be adjusted dynamically when the Manager inserts them into TS. The system could adapt the task size at runtime to accommodate significant changes in computational capacity. For example, when so many low-performance devices join (favoring smaller tasks) or when high-performance devices dominate (favoring larger tasks to reduce I/O overhead). However, in this paper, to make it simple we assume a fixed task size, which is determined in advance and remains unchanged during execution.

\textbf{Task Size}: The size of a task is influenced by multiple factors, such as the memory required to store parameters and the number of multiplications/additions involved. In particular, the loss task typically involve more complex computations and are better to be assigned a proportionally larger size.

Task partitioning is performed by reducing the input and output sizes of a task under the same type. For example, a forward task with input dimension $m$ and output dimension $n$ can be split into \textbf{four} smaller forward tasks, each corresponding to one formed by dividing the input and output in half. Specifically, these sub-tasks cover: (first $m/2$ inputs \& first $n/2$ outputs), (first $m/2$ \& last $n/2$), (last $m/2$ \& first $n/2$), and (last $m/2$ \& last $n/2$). Since tasks are represented as descriptions, the Handler can independently retrieve the necessary weights and biases from TS during execution.

Backward tasks follow the same logic, as they also depend on both input and output dimensionality to retrieve the weights/biases. For other task types, splitting is more straightforward:

\begin{itemize}
    \item An activation task operating on an $m$-dimensional vector is split into two tasks, each handling $m/2$ elements.
    \item An $m$-dimensional update tasks are similarly split into two tasks, each updating $m/2$ parameters.
    \item An $m$-dimensional loss tasks are divided into two smaller tasks, each responsible for computing loss over $m/2$ outputs. The Handler automatically retrieves the corresponding ground truth labels during processing.
\end{itemize}

After generating the prototype tasks for each NN layer, the Manager process them according to the predefined task size and inserts the resulting task descriptions into TS, where they await propagating by the Manager.

\subsection{Task Dispatching}

Following the design of ACAN, the Manager retrieves from TS both the data required for the current stage of NN training and a pouch of partitioned tasks (up to the predefined pouch size). These tasks are then inserted into TS and remain available for processing for a duration defined by the timeout.

Each Handler continuously attempts to process tasks by taking one available task from TS under the key $task$. If the preconditions are met, the Handler performs the computation and writes the results back to the corresponding TS entries. Additionally, it marks the associated keys for task completion, enabling the Manager to track progress. The above process is repeated continuously by all Handlers.

After the timeout, the Manager checks the status of all previously issued tasks. Based on how many tasks were completed, it adjusts the timeout to adapt to the current system computational power. The Manager then collects all uncompleted tasks (and possibly new tasks from TS if it is not enough to give a pouch) and re-issues them. If a particular data item or network layer has been fully processed, the Manager updates the relevant TS entries as a checkpoint, so as to reset some flags.

\subsection{Conflict Resolution}

Once a task is published by the Manager, the Handler that takes this task temporarily gains the right to access the related variables. For example, in a forward task, the Handler reads the input data, weights, and biases, and writes the resulting output.

However, due to ACAN's handler-agnostic nature, a Handler may take a task but fail to complete it before the timeout. From the Manager's perspective, the task is incomplete and will be reissued. Meanwhile, the original Handler may still proceed, potentially leading to duplicate execution of the same task.

For tasks that do not write the read variables, such as forward (it does not need to overwrite the inputs, weights and biases), activation, backward, and loss tasks, duplicate execution poses no correctness issue. As long as each execution is valid (a property ensured by the blockchain-like consistency of TS), redundant computation is okay.

In contrast, update tasks involve overwriting the read variables. Redundant execution in this case may result in unintended cumulative updates. Checking task ID for completion and eliminate concurrent updates can be easily done following the same sliding-window discipline in TCP/IP protocols.  Once all update tasks for a given layer are complete, the Manager can safely overwrite them.

\section{Experiments}

Before orchestrating a large ACAN cloud, we simulate our method using multi-threading on a single machine. The target model is a neural network comprising two linear layers, with the first layer sized $N\times N$ and the second $N\times 1$, where $N=4^4$. Each Handler is constrained to a maximum capacity of $4^4$, which is also the upper bound for task size in ACAN. Consequently, the overall model is about 256 times larger than what a single device (Handler) can accommodate, highlighting the necessity of task-level parallelism.

Instead of adopting the Gossip protocol or implementing ACAN on an actual blockchain infrastructure, we simulate the tuple space with a daemon thread which maintains a Python dictionary. For simplicity, we also assume that all Handler threads behave correctly and do not return malicious results.

Our simulation involves \textbf{one Manager thread and four Handler threads}, all of which may crash during execution. The daemon thread continuously monitors the system and revives failed Manager thread using the latest checkpoint. Although ACAN is designed to be Handler-agnostic and does not require active fault recovery for Handlers, in our simulation we still recreate crashed Handler threads. Instead, to emulate fluctuating computational resources, we dynamically vary the processing speed of Handler threads during runtime. To highlight the robustness of our approach, we fix both the task size ($4^4$) and the pouch size (100) throughout the experiment and only adjust the timeout.

We conduct three experiments in total:

\begin{itemize}
    \item Experiment 1 serves as a basic feasibility check. Both the Manager and Handlers remain stable, and the processing speed is fixed. We perform a regression task to demonstrate that the NN can be successfully trained under our method.
    \item Experiment 2 introduces dynamic changes in Handler processing speeds while still assuming no crashes. This allows us to evaluate whether ACAN's adaptive timeout mechanism can effectively track performance shifts across Handlers.
    \item Experiment 3 builds upon Experiment 2 by introducing probabilistic failures for both the Manager and Handlers. We analyze regression loss and timeout dynamics to assess the method's resilience under fault-prone conditions.
\end{itemize}

\subsection{Feasibility Test}

To construct a learnable target function, we randomly generate a set of parameters that define a mapping from a $4^4$-dimensional input to a 1-dimensional output. Based on this function, we synthesize 100 data points and train the neural network using the Mean Squared Error (MSE) loss. The goal is to verify whether the proposed method enables the network to train properly.

Given that the primary focus of this experiment is to validate the operational feasibility rather than assess generalization performance, we do not generate a large dataset or include a separate test set. For the same reason, we forego mini-batch training and instead use one data point at a time (i.e., stochastic gradient descent with batch size = 1).

Figure.\ref{fig:ftest} below shows the training loss curve over 2 epochs.

\begin{figure}
    \centering
    \includegraphics[width=0.9\linewidth]{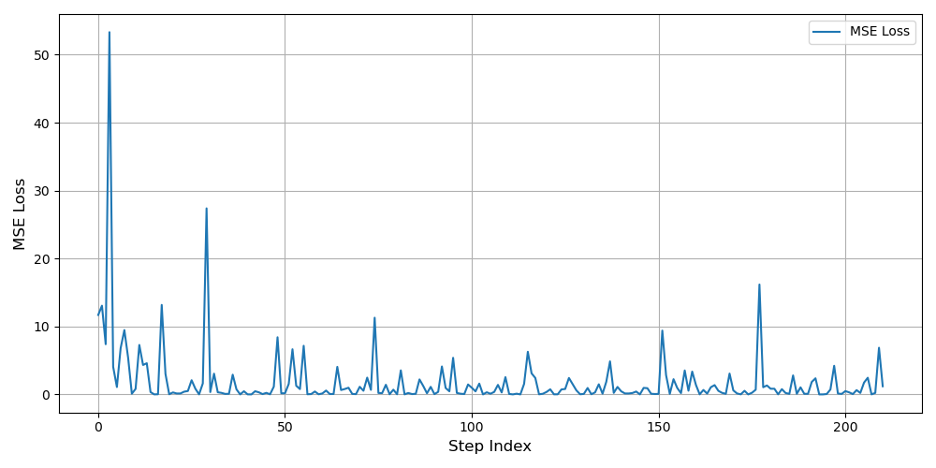}
    \caption{The MSE loss curve in the feasibility test.}
    \label{fig:ftest}
\end{figure}

Observing an effective decrease in the loss function is enough to show that the method is working properly.

\subsection{Adaptability Test}

We set 3 different performance levels for Handlers, with a processing speed ratio of 1:5:10. During training, each Handler has a 100\% chance of random changing every 5 seconds. After processing a task, each Handler will sleep for a period of time based on the processing speed and the size of the task being processed.

\begin{figure}
    \centering
    \includegraphics[width=0.9\linewidth]{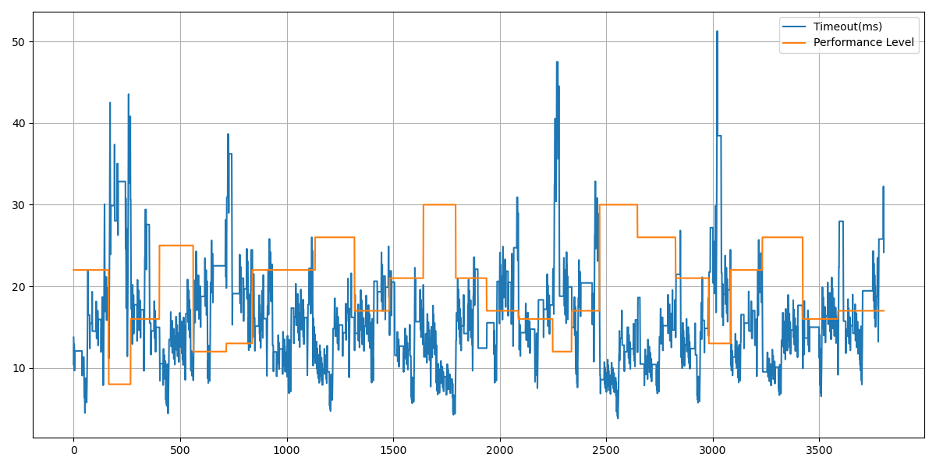}
    \caption{The relation between the timeout and the computational power in the adaptability test.}
    \label{fig:atest}
\end{figure}

We processed a total of 20 data (about 3,500 pouches) and obtained Figure.\ref{fig:atest} of Handler computing power and ACAN timeout, which showed an obvious inverse proportional relationship, indicating that the system can use lower timeout to speed up processing when computing power is high, and vice versa.

\subsection{Robustness Test}

Based on the adaptability test, we make the Manager 100\% crash every 5 seconds, all Handlers 100\% crash every 5 seconds, and 100\% change the processing speed every 5 seconds. We also tested on 20 data and get the following curve.

\begin{figure}
    \centering
    \includegraphics[width=0.9\linewidth]{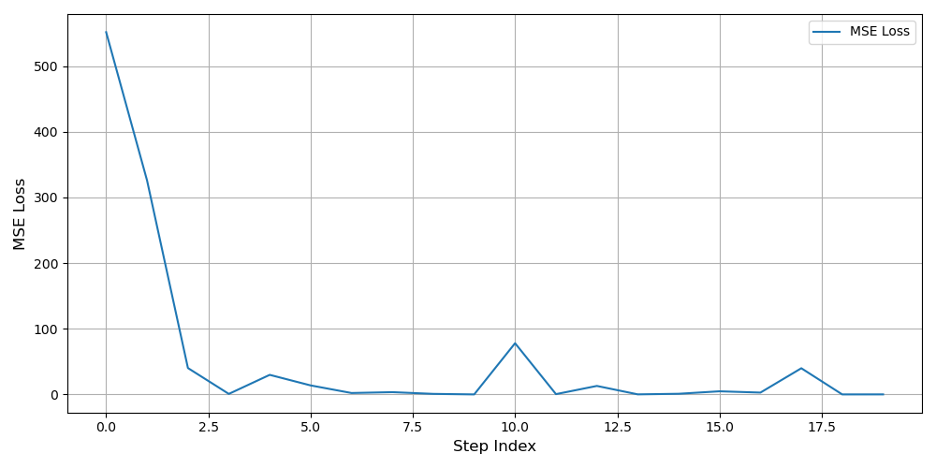}
    \caption{The MSE loss curve in the robustness test.}
    \label{fig:rtest-1}
\end{figure}

\begin{figure}
    \centering
    \includegraphics[width=0.9\linewidth]{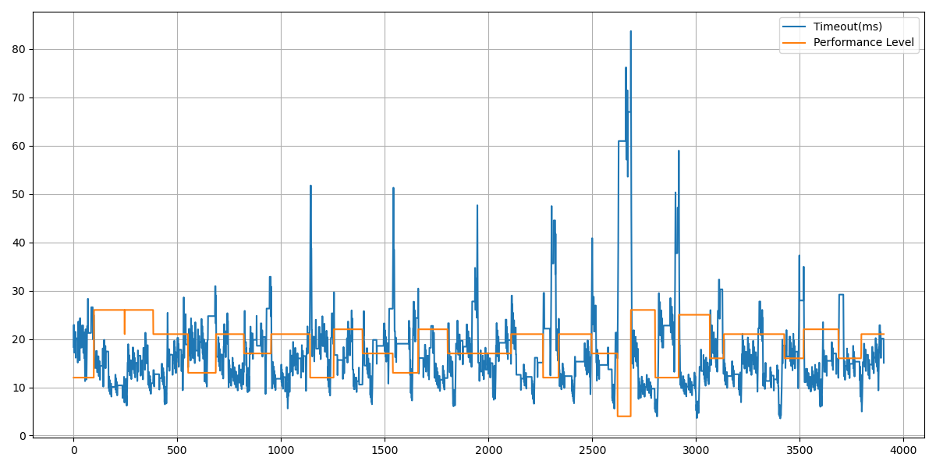}
    \caption{The relation between the timeout and the computational power in the robustness test.}
    \label{fig:rtest-2}
\end{figure}

The loss curve in Figure.\ref{fig:rtest-1} still shows that the method is working properly, while the performance curve in Figure.\ref{fig:rtest-2} shows that the overall relationship between timeout and performance level is still inversely proportional, indicating that this method can still show stability under high crash conditions. The number of pouches to be processed increases due to the extremely high crash probability.

\section{Significance}

The scale of neural network training continues to grow rapidly due to intense AI/ML competition. The reconfigurable von Neumann multiprocessor architecture promises to deliver sustainable performance and reliability regardless of scale. The reported experiments used extreme cases to prove the feasibility of these claims.

\section{Design Tradeoffs}

The reported experiments used a dictionary to simulate the ACAN Tuple Space. It has identical data retrieval semantics, without the program-to-program synchronization semantics. Therefore, the performance may be better due to the absence of delays.

Using a tuple space for indirect program-to-program communication will incur higher communication overhead relative to direct program-to-program communication (nearly 2x). However, the savings in minimized synchronization overheads and checkpoint delays with the custom cloud's natural parallelization effects, the proposed method can still out-perform direct program-program communication counterparts even without checkpoints. The energy savings would allow training of even larger models.

The ACAN Tuple Space service is implemented in reality as an extension to operating systems communication stack for each participating node. Unless all nodes fail at the same time, the application can still continue. However, data corruption during transmission or storage requires special application programming to ensure the quality of data delivery. These small probability errors are typically negligible in scientific simulations and NN training applications.

\section{Conclusion}

The reported study proves the feasibility of a reconfigurable multiprocessor architecture for NN workflows. The next step is orchestrating a large custom ACAN for targeted applications. In particular, experiments with Tenstorrent processors can possibly shed more light on the proposed programming paradigm. 

The proposed ACAN concept and protocols \cite{ShiPatent16} depart from the traditional client-server programming paradigm in its complete program and data decoupling features. Rather than relying on compilers to map applications onto multiprocessors, the proposed framework enables runtime formation of computational graphs dynamically. It is also possible to compliment compiler generated programs to overcome the scaling challenges without checkpoints.

A salient feature of the proposed framework is "unlimited scalability" due to the resolution of the SPF challenge. Without SPF threats, all applications can scale indefinitely, limited only by available resources. Although security was not discussed in this report, the scalability feature can compensate for any security measure overhead by simply expanding the size of the infrastructure.

ACAN further restricts the threat vectors within the application boundary. Therefore, the proposed framework is suitable for all mission-critical applications.

\section{Acknowledgment}
The reported work is supported in part by NSF Chameleon resource grants \#CH-817746 and \#CHI-251439, Temple University Technology Transfer Office, Temple AGI Team and the SMC Labs. LLC.

\bibliographystyle{splncs04}
\bibliography{samplepaper}
\end{document}